\documentclass[prl,twocolumn,floatfix]{revtex4-1}
\usepackage{amssymb}
\usepackage{graphicx}
\usepackage{caption}
\usepackage{subcaption}
\usepackage{multirow}
\usepackage[table,xcdraw]{xcolor}
\usepackage{soul}
\usepackage{kantlipsum}
\usepackage{amsmath}
\usepackage{amsfonts}
\usepackage{amssymb}
\usepackage{dcolumn} 
\usepackage{braket}
\usepackage{booktabs}
\usepackage{multirow}
\usepackage[table]{xcolor}
\usepackage{dsfont}
\usepackage{times}
\usepackage{epsfig}
\usepackage{xcolor}
\usepackage{tikz}
\usepackage{float}
\usepackage{color}

\usepackage{siunitx}

\usepackage[english]{babel}

\begin{document}
\title[]{Ultra-thin, entirely flat, Umklapp lenses}
\author{Gregory~J.~Chaplain$^{1}$ and Richard~V.~Craster$^{1}$}
\affiliation{$^1$ Department of Mathematics, Imperial College London, London SW7 2AZ, UK }

\begin{abstract}
We design ultra-thin, entirely flat, dielectric lenses using crystal momentum transfer, so-called Umklapp processes, achieving the required wave control for a new mechanism of flat lensing; physically, these lenses take advantage of abrupt changes in the periodicity of a structured line array so there is an overlap between the first Brillouin zone of one medium with the second Brillouin zone of the other. At the interface between regions of different periodicity, surface, array guided, waves hybridise into reversed propagating beams directed into the  material exterior to the array. This control, and redirection, of waves then enables the  device to operate as a Pendry-Veselago lens that is one unit cell in width, with no need for an explicit negative refractive index. Simulations using an array embedded in a slab of silicon nitride ($\text{Si}_3\text{N}_4$) in air, operating at visible wavelengths between $420 - 500\si{\tera\hertz}$ demonstrate the effect.
\end{abstract}
\maketitle

\begin{figure}[t!]
    \centering
    \includegraphics[width = 0.4\textwidth]{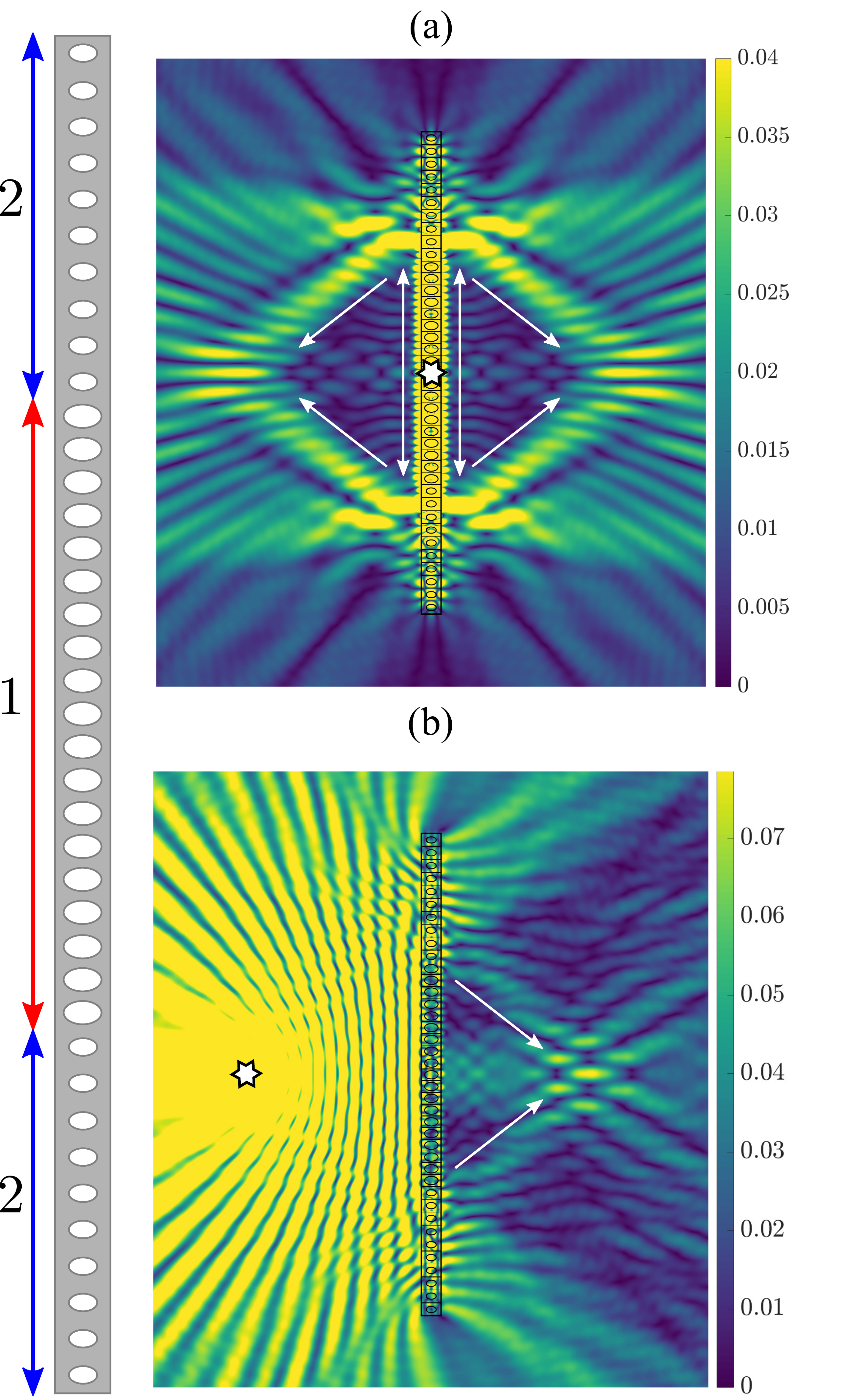}
    \caption{Dielectric thin flat lens: Electric field, $\vert E\vert$, shown in arbitrary units for line source excitation, marked at white star operating at frequency $484\si{\tera\hertz}$. (a)  Array centred source, giving reversed conversion via the Umklapp mechanism. (b) Flat lensing by source placed at $-8\lambda$, producing image at opposite focal spot. The array is shown on the left: device length is $L = 12.3\si{\micro\meter}$, with width $w = 500\si{\nano\meter}$.}
    \label{fig:Fields}
\end{figure}

\noindent \textbf{\textit{Introduction:}} Inspired by the ability to create flat lenses, such as the Pendry-Veselago lens \cite{Pendry2000}, there has been a drive towards developing flat optical devices to manipulate light \cite{Yu2014,lalanne2017metalenses}. Advances in metasurface-based technologies have paved the way for these planar devices, with realisations in a variety of settings, operating by several different modalities. These operational regimes include plasmonic waveguide structures \cite{Verhagen2010}, i.e. bulk materials composed of alternating metal-dielectric layers \cite{Xu2013}, or all dielectric inhomogeneous layered lens antennas \cite{AlNuaimi2014DesignOI}. Alternate flat lens devices focus on controlling abrupt phase changes endowed to incident wavefronts upon transmission \cite{Khorasaninejad2016,Genevet2017}, taking advantage of gradient index structures \cite{Jin2019}, or by manufacturing protruding subwavelength elements, enabling generalised refractive laws to be observed \cite{Yu2014}. Other materials, based on two dimensional photonic crystals have achieved negative refraction, without the need of a negative index \cite{Luo2011}. Unlike conventional metalens or diffractive lenses \cite{banerji2019imaging}, here we present a new, novel, flat dielectric lens antenna (DLA) device based on simple \textit{singly} periodic structures, with positive refractive index operating over the frequency range $420-500\si{\tera\hertz}$. These devices operate by promoting Umklapp scattering \cite{Peierls_Thesis} at a designed region, and offer a new modality of flat lensing, as shown in Fig.~\ref{fig:Fields}.

Recent design paradigms for adiabatically graded arrays have resulted in a remarkable level of wave control, and phenomena being observed, in multiple disciplines within wave physics stretching from elastic vibrations and acoustics through to electromagnetism. The inspiration of many such designs stem from the rainbow trapping effect, originating in electromagnetism \cite{Tsakmakidis2007}, whereby the speed and phase of localised array guided modes is manipulated by graded geometrical changes of the array components. Elementary resonant, often sub-wavelength, devices have been proposed \cite{Chaplain2019,colombi16a}, and built \cite{Colombi2017},  for elastic media enabling trapping, and mode conversion transferring energy from surface to body waves,  effects for array guided surface states for applications to energy harvesting \cite{DePonti2019} from vibration. Recently, a reversed conversion phenomena which emulates negative refraction by a line array \cite{Chaplain2019Flat} has been developed using a counter-intuitive effect hybridising both trapping and conversion; this relies upon band crossings and phase-matching all within the first Brillouin zone and uses adiabatic grading of an array all in the setting of elastic waves.    

In this article we propose a much more versatile design paradigm employing Umklapp "flip-over" processes to achieve flat lensing; it is remarkable that this concept from a disjoint area of physics, Umklapp processes primarily arise in thermal transport, can be adapted to optics. The operation rests on the segregation of the device into structured regions of different periodicities, thereby creating two different Brillouin zones in reciprocal space, and the subsequent analysis of the corresponding dispersion curves. Contrary to the adiabatically graded arrays considered in  \cite{Chaplain2019Flat}, there is no grading between the regions but instead an abrupt change in the array periodicity; undesirable scattering of the field at such an interface is anticipated, but by  carefully engineering the design we can recapture the scattered field by promoting Umklapp processes thereby providing a remarkably simple way to achieve flat lensing, without the need for exotic materials or inhomogeneities.

\begin{figure}
    \centering
    \includegraphics[width = 0.4\textwidth]{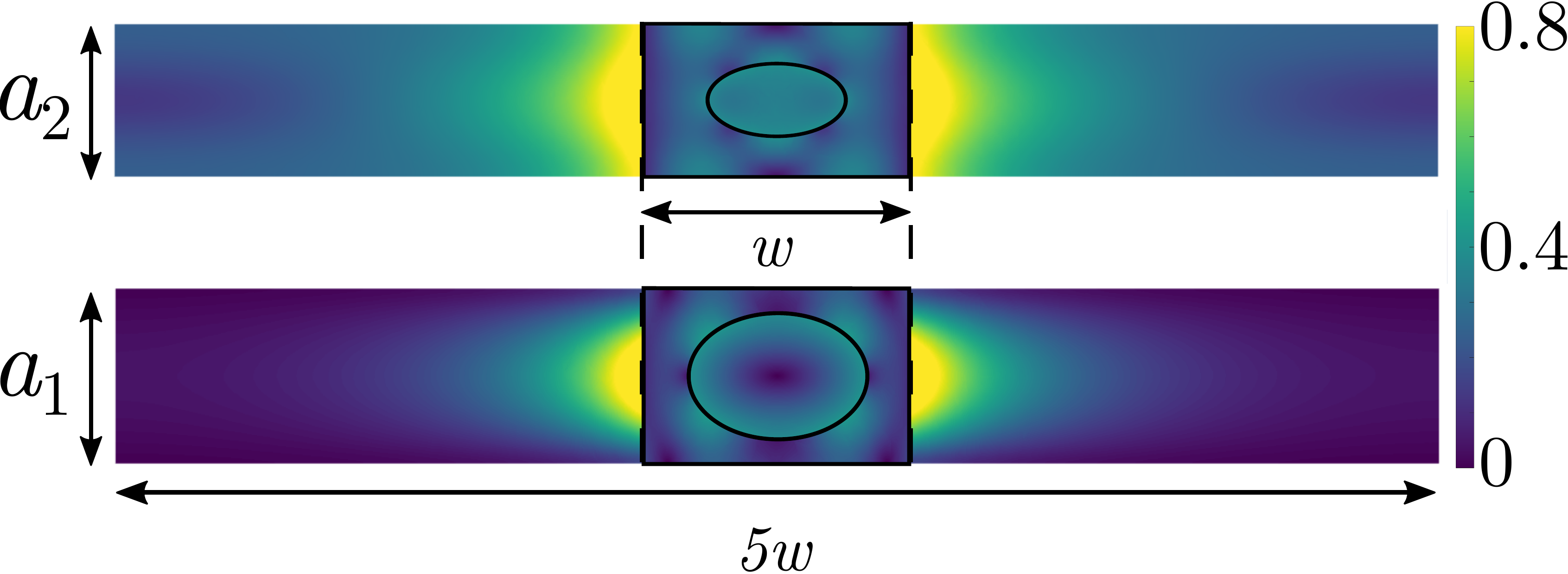}
    \caption{Typical electric fields, of similar mode symmetry, as $\vert E\vert$ for each region. The elliptical inclusions have semi-major and minor axes $r_{a}$ and $r_{b}$ respectively, detailed in Table.~\ref{Tab:param}. }
    \label{fig:Schem}
\end{figure}

We consider transverse electric (TE) \cite{Naqvi1990} polarised time harmonic fields satisfying the scalar equation 
\begin{equation}
    \nabla \cdotp (\tilde{\mu}_{r}^{-1}\nabla E) + \epsilon_{r}\kappa_{0}^2E = 0,
\end{equation}
where $\tilde{\mu}_{r}$ and $\epsilon_{r}$ are the relative permeability and permittivity respectively and $E$ the out-of-plane electric field. We focus on localised surface electromagnetic waves confined to a dielectric slab in air, chosing silicon nitride $\text{Si}_3\text{N}_4$ as the dielectric as it is widely used in applications due to its high relative permeability ($\epsilon_{r} = 9.7$), ubiquity in integrated circuitry \cite{pierson1999handbook} and transparency over the visible range in nanoengineered structures both with low contrast ($n \approx 2$) and loss \cite{zhan2016low}. The supported surface states manipulated throughout are neither surface plasmon polaritons, or their spoof counterparts \cite{Ritchie1957,Pendry2004}, since we do not deal with opposing signs of permittivities or structured metallic interfaces. Indeed the proposed structuring takes place within the device, leaving its edges completely flat (Fig.~\ref{fig:Schem}). As such we are free to model these devices efficiently in air operating in the terahertz frequency range; the corresponding device dimensions are achievable in waveguide technologies \cite{munoz2017silicon,Ghadimi2018}. We do not exploit resonance effects, and so the imaging is not subwavelength, yet the proposed DLA devices, with their resilience to surface imperfections \cite{lo1993antenna,wang2006design}, are ideal for demonstrating the efficacy of Umklapp scattering.

Our analysis is, theoretically, independent of the material and device size, and rests upon exploiting the dispersion curves and isofrequency contours of the different periodicities within the device, as detailed by the design methodology. Using this, we then outline how to harness this promoted scattering effect to generate a new mechanism for flat lensing, and investigate the effects of losses in the device. Suitable scaling to other materials, and sizes, permits the tunability of the frequency bandwidth, leading to a new operational capacity for dielectric substrates. 

\noindent \textbf{\textit{Design Methods:}}
\label{sec:design}
When analysing periodic media it is naturally convenient to display the dispersion diagrams of such materials within the irreducible Brillouin zone (IBZ) \cite{brillouin2003wave}. For materials exhibiting different periodic regions capable of supporting surface waves, phenomena occurring at frequencies corresponding to wavevectors lying outside the first Brillouin zone (BZ) are therefore missed. Conventionally, we concentrate upon modes below the dispersionless light line of free space waves, that is, within the first BZ; Umklapp processes, arising due to the transfer of crystal momentum from higher BZs, are thereby ignored. We  demonstrate that there are advantages in using these processes by considering two regions of differing periodicities within the same dielectric array, see Fig. 1, such that there is an overlap between the first BZ of one periodic region and the second BZ of the other. At the abrupt transition between these regions Umklapp scattering is dominant and reversed conversion can be achieved and utilised for flat lensing.

The Umklapp mechanism, first hypothesised by Peierls in 1929 \cite{Peierls_Thesis} is conventionally used to describe thermal transport and resistivity of metals at high temperatures, and is now prevalent in the quantum theory of transport \cite{peierls1996quantum,taylor2002quantum}. It is a manifestation of the transfer of crystal momentum within the system, and exploits the fact that in periodic media wavevectors are defined up to a reciprocal lattice vector, $\boldsymbol{G} \equiv 2\pi/a$, with $a$ being the periodicity of the unit cell. Thus in the case of scattering two initial wavevectors, say,  $\boldsymbol{\kappa_{1}}, \boldsymbol{\kappa_{2}}$ then if the resultant $\boldsymbol{\kappa_{3}}$ lies beyond the first Brillouin zone it experiences the Umklapp, or `flip-over' mechanism via crystal momentum transfer \cite{Maznev2014}. Two types of scattering processes are defined: normal (N-processes) and Umklapp (U-processes) through
\begin{equation}
    \boldsymbol{\kappa}_{1} + \boldsymbol{\kappa}_{2} - \boldsymbol{\kappa}_{3} = \begin{cases}
    \boldsymbol{0} & \text{N-process}, \\
    \boldsymbol{G} & \text{U-process}.
    \end{cases}
    \label{eq:umklapp}
\end{equation}
This mechanism has recently been observed to cause excess resistivity in graphene \cite{Wallbank2019} and used to explain the coupling of acoustic and optical branches in crystals \cite{Bauer1957}. We apply it, for the first time, in a flat lensing scenario, circumnavigating any ambiguities considered with Umklapp scattering \cite{Maznev2014}, by adopting the conventional description in equation \eqref{eq:umklapp}, analysing U-processes for scattered wavevectors lying outside the first BZ; this is a natural way to distinguish between N- and U-processes and is critical in this design process. 

For the application of the Umklapp mechanism to localised surface electromagnetic (EM) waves for flat lensing, we partition a dielectric slab into two regions of differing periodicities, see Fig. 1. To set the context, we first analyse  a perfectly periodic medium of consisting of a slab of $\text{Si}_3\text{N}_4$ in air structured with a periodic array of elliptical inclusions with pitch $a_1$, with geometrical parameters given as region 1 in Table~\ref{Tab:param} as shown in Fig~\ref{fig:Schem}. Then consider the same dielectric material but with the array having a larger periodicity, $a_2$,  and different inclusion size (region 2 in Table~\ref{Tab:param}). The two regions are defined with unit cells of length $a_{1}, a_{2}$ respectively, such that $a_{2} > a_{1}$. Consequently the first BZ of region 2 is smaller than that of the region 1. Thus, the edges of the first BZ for the regions, $X_{i} \equiv \pi/a_{i}$, satisfy $X_{2} < X_{1}$ and are offset from each other. The dispersion curves within this overlapping region are calculated using the FEM software Comsol multiphysics and shown in Fig.~\ref{fig:Disp}. The frequencies where overlap between the dispersion curves of similar mode shapes are where U-processes take effect efficiently; the excited mode in region 1 must be able to excite a mode that exists in the region 2, otherwise an effective hard boundary is reached at the transition region, resulting in undesirable scattering; typical fields showing similar modal symmetries in the two regions are shown in Fig.~\ref{fig:Schem}.  Inspecting dispersion branches with opposing signs of group velocity can also give insight into where this reversal takes place \cite{Maznev2014}. This effect is achieved for several of the overlapping bands shown in Fig~\ref{fig:Disp}, ranging between $420-500\si{\tera\hertz}$, and therefore this offers broadband performance. A further example of the reversed conversion effect is shown in supplemental Fig.~S1.
\begin{table}[t]
\begin{minipage}[b]{0.55\linewidth}
\centering
 \begin{tabular}{||c c c||} 
 \hline
  & Region 1 & Region 2 \\ [0.5ex] 
 \hline\hline
 $a$ & 300\si{\nano\meter}  & 330\si{\nano\meter}  \\ 
 \hline
 $w$ & 500\si{\nano\meter} & 500\si{\nano\meter}  \\
 \hline
 $r_a$ & 170\si{\nano\meter} & 80\si{\nano\meter}  \\
 \hline
 $r_b$ & 110\si{\nano\meter} & 130\si{\nano\meter}  \\ [1ex] 
 \hline
\end{tabular}
\caption{Parameters within dielectric slab; region 1(2) coloured  red(blue) in Fig~\ref{fig:Iso}(a,b).}
\label{Tab:param}
\end{minipage}\hfill
\begin{minipage}[b]{0.44\linewidth}
\centering
\includegraphics[scale=0.3]{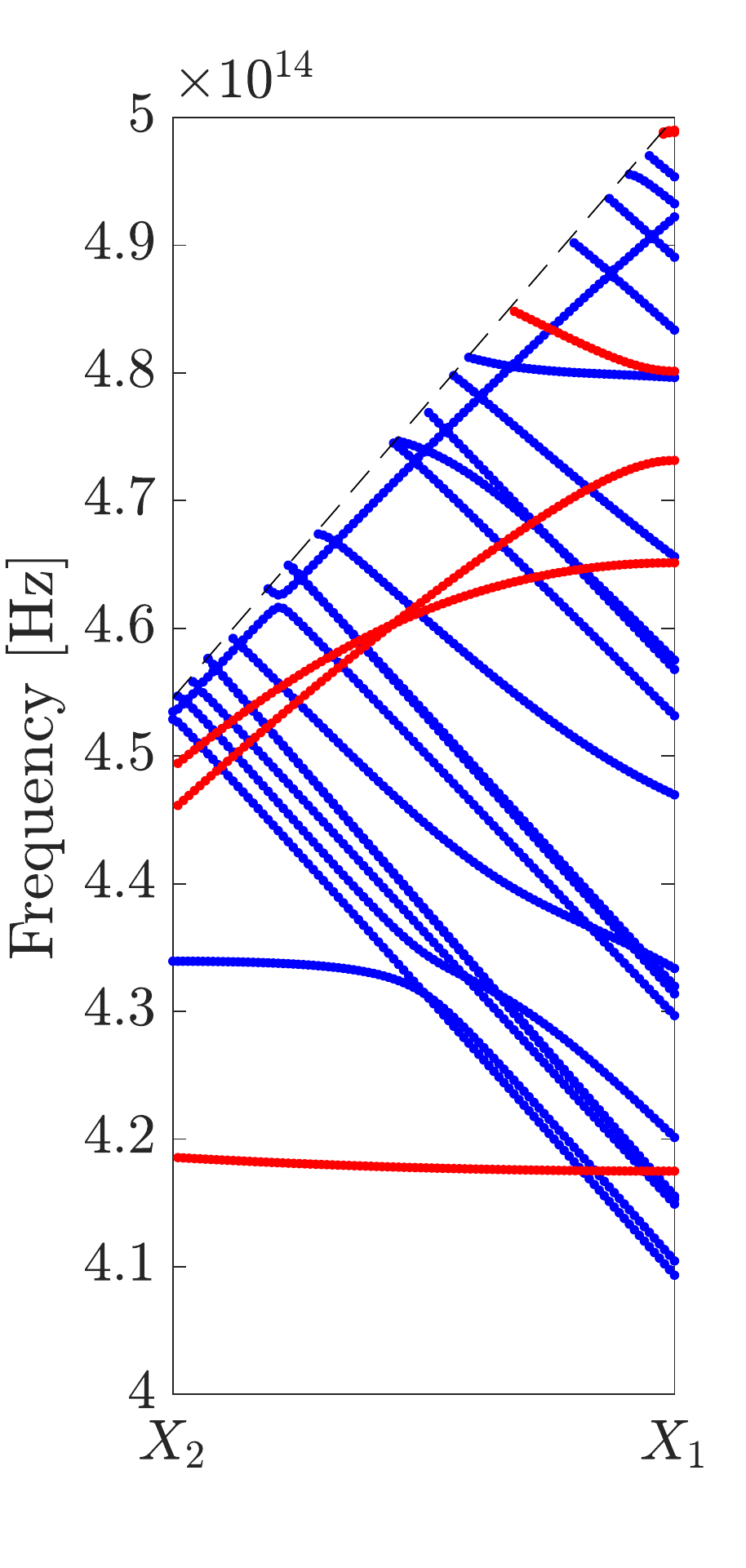}
\end{minipage}
\captionof{figure}{Dispersion curves for region 1(2) in red(blue), plotted from $X_{2}$ to $X_{1}$ i.e. in the second BZ of region 2 whilst in the first of region 1. By matching mode shapes incident modes (red) can scatter into modes in region 2 (blue) via U-processes. Full dispersion curves for each region shown in supplemental Fig.~S2.}
\label{fig:Disp}
\end{table}

\begin{figure}
    \centering
    \includegraphics[width = 0.45\textwidth]{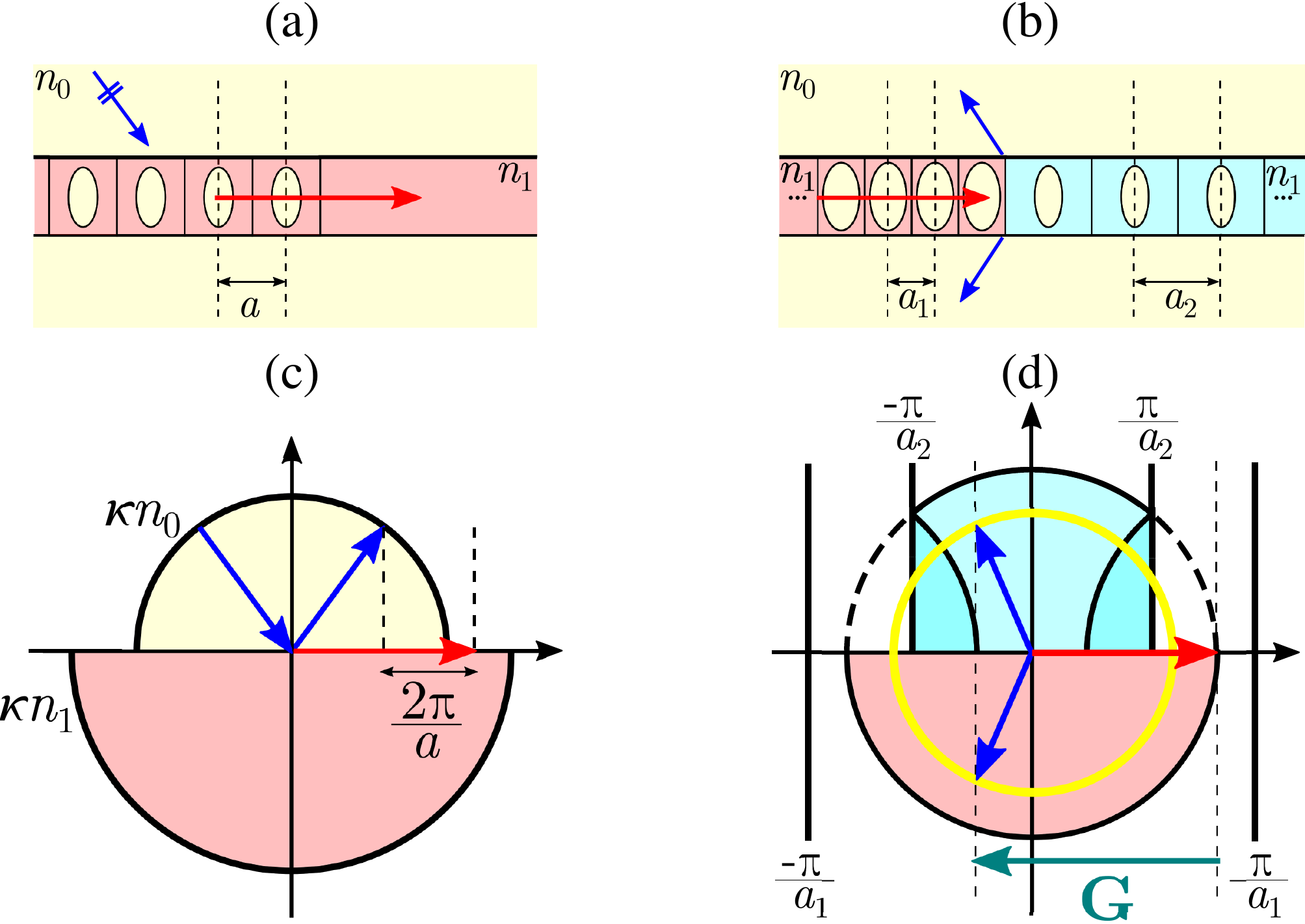}
    \caption{Mode coupling and conversion: (a,c) Conventional mode coupling and (b,d) Umklapp coupling.   (a) incident wave (blue arrow) couples to array guided mode (red arrow) through periodicity of the array. (c) Wavevector interpretation for media of differing refractive indices. (b) Array guided wave (red arrow) reverse converts to propagating waves (blue arrows) upon reaching different periodicity region. (d) Half-plane isofrequency contours for same refractive index media, $n_{1}$, now partitioned into regions of differing periodicity. The wave transitions into second BZ in region with larger physical unit cell (blue region), experiencing Umklapp scattering via crystal momentum transfer allowing hybridisation with exterior $n_{0}$ medium (yellow circle).}
    \label{fig:Iso}
\end{figure}

We now consider the implications of structuring a finite slab of $\text{Si}_{3}\text{N}_{4}$ into two distinct regions, one with the periodicity and unit cell structure of region 1, in Table~\ref{Tab:param}, that then transitions abruptly to incorporate the parameters associated with region 2. Conversion of the array guided mode occurs into a beam in the exterior medium that is directed backward; the angle of the reversed conversion is predicted by inspecting the isofrequency contours of the two regions. Shown in Fig.~\ref{fig:Iso} are simplistic isofrequency contours for the respective components of the dielectric slab; we show for clarity the isofrequency contour of each medium  as a circle, acknowledging that it in fact corresponds to a point on the $\kappa_{x}$ axis when projecting the dispersion curves into wavevector space. Exploiting the coupling at a sharp change in periodicity is inspired by conventional mode coupling theory \cite{chuang2012physics}; incident radiation of wavevector $\boldsymbol{\kappa}$ in a media with refractive index $n_{0}$, is coupled to a waveguide mode of vector $\boldsymbol{\kappa}_{wg}$ through phase matching by incorporating the periodicity, such that
\begin{equation}
    \boldsymbol{\kappa}_{wg} = \boldsymbol{\kappa}n_{0}\sin\theta + \Lambda,
    \label{eq:mode}
\end{equation}
where, for integer $m$, $\Lambda = 2m\pi/a$. This is shown in Fig.~\ref{fig:Iso}(a,c). Using this mode coupling picture we can interpret the hybridisation mechanism from the `contours' of the medium composed of differing periodicities. 

In Fig.~1(a), the array is excited with a source to create a surface wave with wavevector in region 1 (red section in Fig.~\ref{fig:Iso}(b)). The corresponding isofrequency contour is in the lower half of the plane in Fig.~\ref{fig:Iso}(d), defined within the first BZ such that $\kappa \in [-\pi/a_{1},\pi/a_{1}] \equiv [-X_{1},X_{1}]$, we then abruptly change the periodicity to that of region 2, marked by the blue region in Fig.~\ref{fig:Iso}(b); the projection of the isofrequency contour for region 2 is in the upper half plane of Fig.~\ref{fig:Iso}(d). At this frequency, the wavevector now lies in the second BZ of region 2, as highlighted in Fig.~\ref{fig:Disp}, since  $X_{2} < \kappa < X_{1}$. The critical observation is that this initial wavevector, marked by the red arrow, experiences a transfer of {\it crystal momentum} via the Umklapp effect, resulting in the translation of a collinear reciprocal lattice vector, $\boldsymbol{G}$, denoted by the folded isofrequency contours. These have the same radii as in region 1 since the material properties have not changed, only the geometry has altered. In Fig.~\ref{fig:Iso} we superimpose the isofrequency contour of the exterior medium that surrounds the array (yellow circle). Phase-matching with this contour gives the resultant scattered wavevector that, in turn, predicts the reversed conversion angles as in Fig.~\ref{fig:Fields}(a). The angle is explicitly predicted from mode coupling analysis by rearranging equation \eqref{eq:mode} to incorporate the effect of the second periodic region; in this setting $\Lambda = 2\pi/a_{2} > \boldsymbol{\kappa}_{wg}$. Thus we can generalise conventional mode coupling techniques for a wave confined to the array arriving at a region with altered periodicity and notably U-processes provide coupling to the first {\it negative} diffractive order of mode coupling theory \cite{chuang2012physics}. By (a)symmetrically introducing abrupt changes in periodicity about a central point, we can image a line source on the array to two focal spots on either side of the slab, as in Fig.~\ref{fig:Fields}(a), and tune the position of these focal points.

\noindent \textbf{\textit{Umklapp Lensing:}}
\label{sec:lens}
Our arguments also generalise for excitation by a source removed from the array. We again use the Umklapp and the reversed conversion mechanisms at interfaces between different regions of periodicity, but now introduce  reciprocity to generate focusing from one side of the slab to the other. An isotropic line source is placed at one focus of Fig.~\ref{fig:Fields}(a), and an image is produced at the other side of the array, showing the device is capable of emulating negative refraction. The focusing response is due to the horizontal wave component igniting the surface guided wave, which propagates along the dielectric array to the regions of altered periodicity, seen in Fig.~\ref{fig:Fields}. Upon reaching these altered regions, Umklapp scattering takes place and the reversed conversion effect acts to refocus this point source on the opposite side of the device, as clearly shown in Fig.~\ref{fig:Fields}(b). Flexibility in the positioning of the focal spots is achieved through altering the position of the transition regions, the relative periodicities and symmetry. Considering the fabrication of such devices atop other structures, as is conventional in dielectric waveguides \cite{munoz2017silicon}, also permits tunability of the focal lengths on either side of the device \cite{Chaplain2019Flat,maradudin2011structured}. 

\begin{figure}[t!]
\centering
    \begin{tabular}{c}
    (a) \\ \includegraphics[width = 0.4\textwidth]{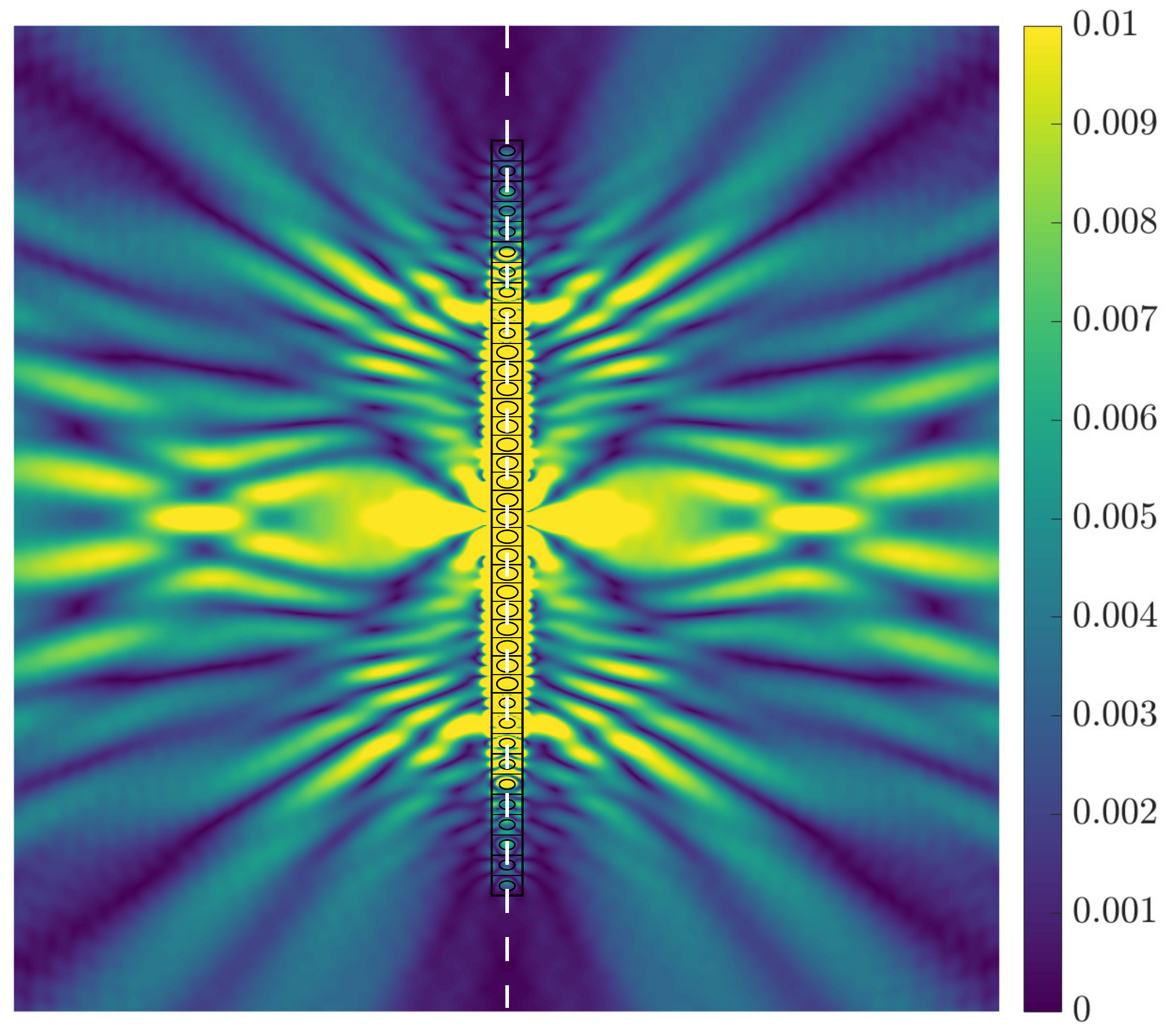} \\ (b) \\ \includegraphics[width = 0.45\textwidth]{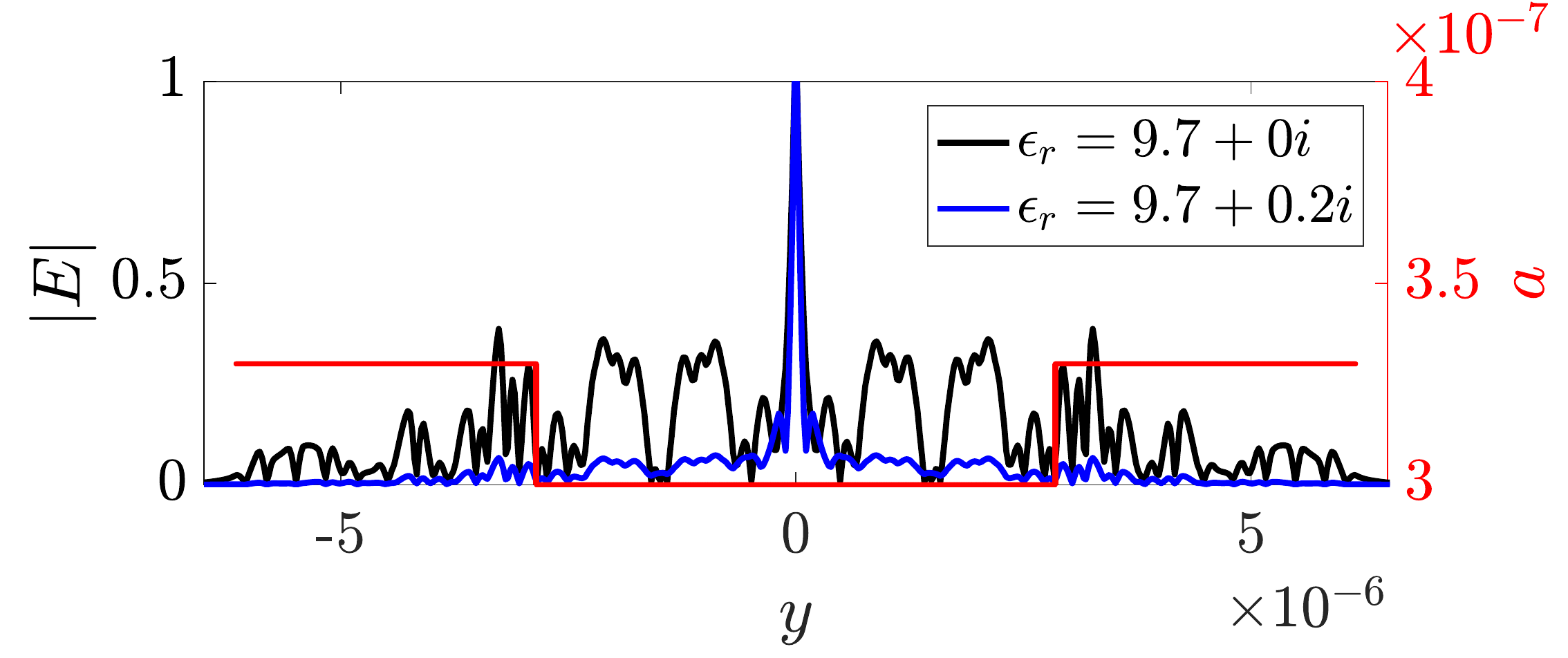}
    \end{tabular}
    \caption{(a) Normalised electric field, $\vert E\vert$, for lossy case with $\epsilon_{r} = 9.7+0.2i$, excited by line source at the centre of the array at frequency $484~\si{\tera\hertz}$. The effect is less pronounced due to weaker propagation along the array. (b) Comparison of normalised electric field norm between lossless (black) and lossy (blue) media, along centre of array (dashed white line in (a)); the local periodicity, $a$, as a function of position is also shown (red). The effect of loss is seen through the reduced field strength along the array.}
    \label{fig:Losses}
\end{figure}

The validity of this concept for devices is further tested by the introduction of losses within the material, through the addition of an imaginary component to the permittivity. An example is shown in Fig.~\ref{fig:Losses}(b), where the field strength along the centre of the devices is compared with, and without, loss. For a lossy material, there is weaker propagation along the array and so the reversed conversion is not as pronounced; this can be mitigated by designing the position of the transition region to lie close to the point of excitation. For decaying surface waves the effect can still be achieved by incorporating the decay length, which can be predicted through homogenisation techniques \cite{Chaplain2019,Chaplain2019Flat}.

\noindent \textbf{\textit{Concluding remarks:}}

We develop the concept of Umklapp lensing by designing an array that requires operation at wavevectors outside the first Brillouin zone, and then manipulating confined surface waves by utilising crystal momentum transfer. The resulting electromagnetic radiation generates focal points, using high permittivity dielectrics that are both entirely flat, and ultra thin: Negative refractive effects can therefore be emulated with by a line array with thicknesses of just one unit cell. 

Considering lossy materials motivates the design of the transition region between differing periodicities, which is key to employing the Umklapp effect. So too is choosing the periodicities in such a way as to achieve maximal overlap of mode shapes within the differing Brillouin zones. At present the proposed devices are purely passive, but there is scope to achieve active components by utilising piezoelectric materials to alter the periodicities and as such, in theoretically demonstrating this new lensing mechanism, we envisage motivation towards experimental verification.

\noindent \textbf{\textit{Funding Information:}} The authors thank the UK EPSRC for their support through Programme Grant EP/L024926/1 and a Research Studentship. R.V.C acknowledges the support of the Leverhulme Trust. 


%

\clearpage
\onecolumngrid
\section{Ultra-thin, entirely flat, Umklapp lenses Supplemental Material}
\setcounter{figure}{0}
\renewcommand{\thefigure}{S\arabic{figure}}

\begin{figure*}[h!]
  \centering
     \includegraphics[width = 0.7\textwidth]{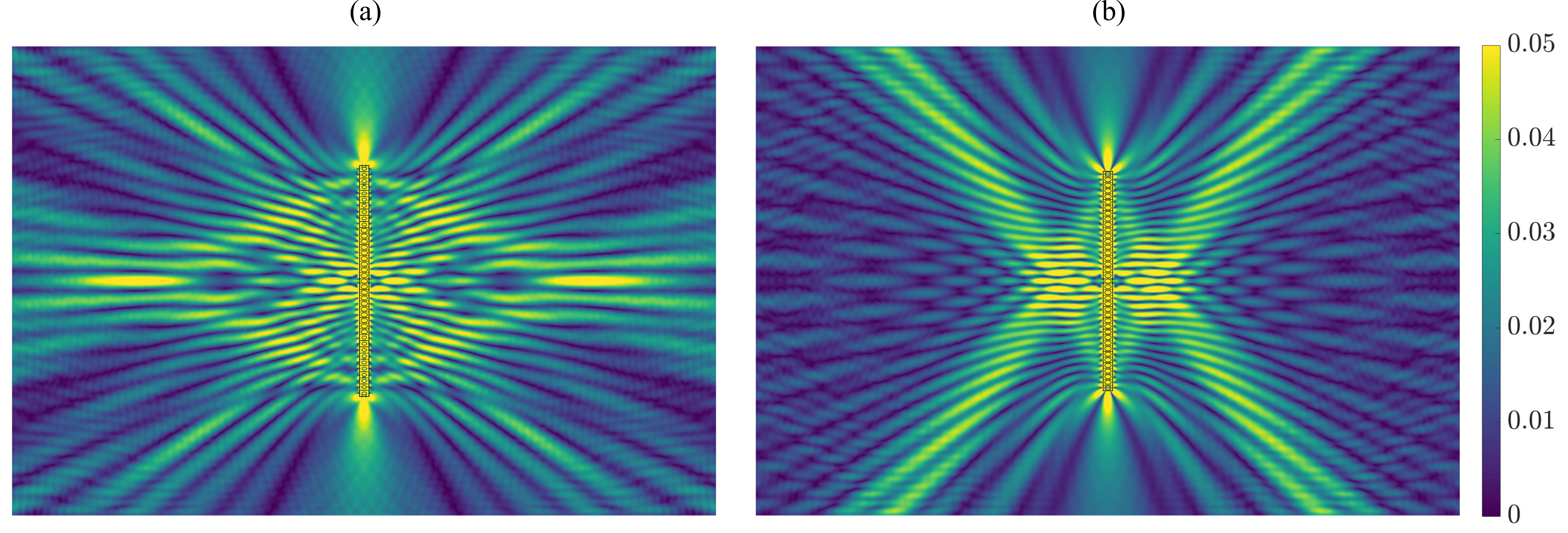}
     \caption{(a) Reversed Conversion effect for excitation frequency of $418\si{\tera\hertz}$, using same array as in Fig.~1(a). (b) Array with no transition region operating at $418\si{\tera\hertz}$, shows no reversed conversion due to lack of U-procceses.}
     \label{fig:LowFreq}
 \end{figure*}

\begin{figure*}[h]
     \centering
     \includegraphics[width = 0.475\textwidth]{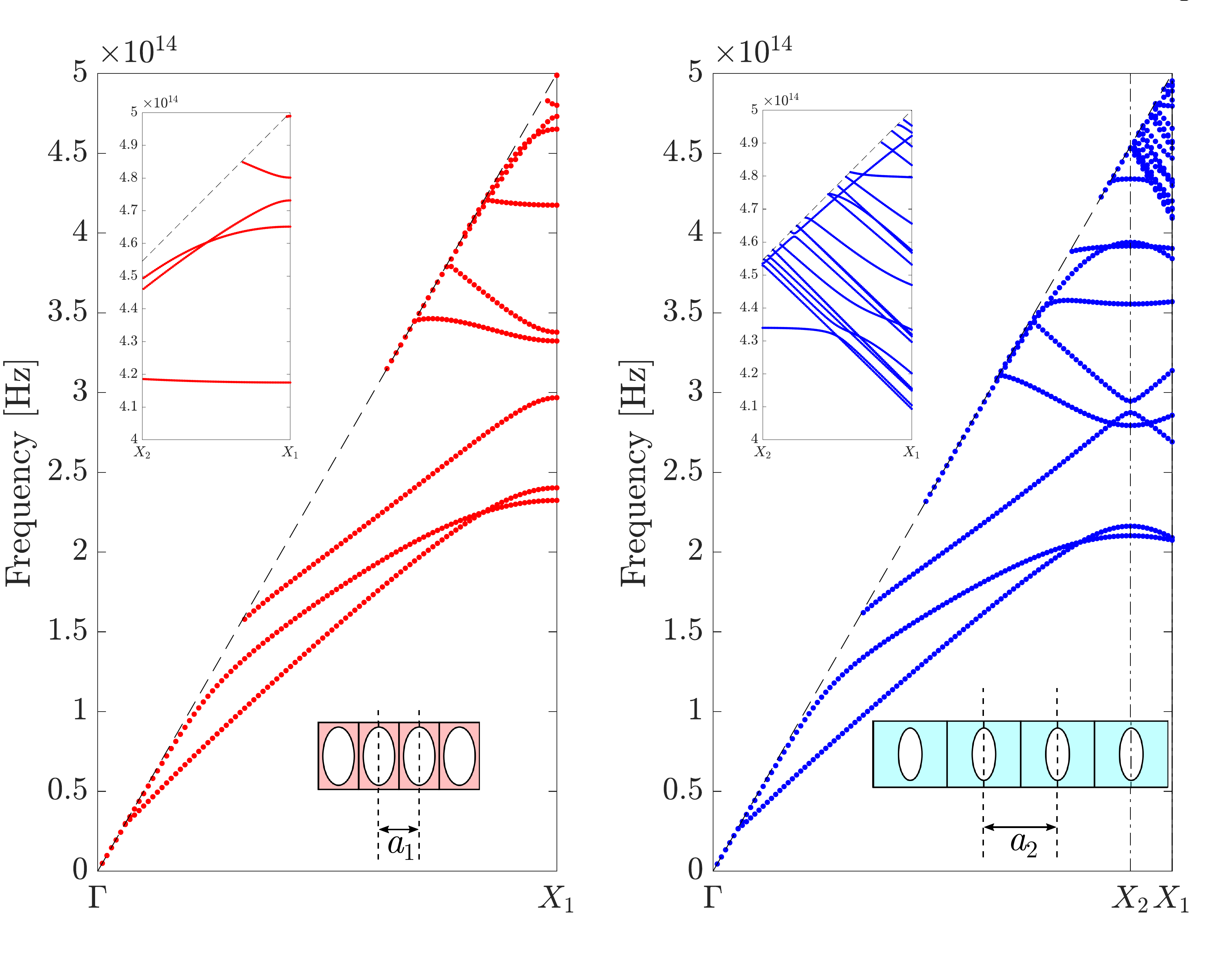}
     \caption{(a) Dispersion curves in first BZ for region 1 in Fig.~3. Insets show zoom of frequencies between $400-500\si{\tera\hertz}$ and unit cell geometry of region 1. (b) Dispersion curves in first BZ of region 2, extending to $X_{1}$ of region 1. Insets show zoom of overlapping region between $400-500\si{\tera\hertz}$ and unit cell geometry for region 2.}
     \label{fig:FullDisp}
 \end{figure*}

 To demonstrate the flat lensing behaviour over a broad range of frequencies, we present another example with the device operating at a frequency of $418\si{\tera\hertz}$. There is an overlap between modes in the first BZ of region 1, capable of exciting modes in the second BZ of region 2, as predicted using Fig.~\ref{fig:FullDisp}. Shown in Fig.~\ref{fig:LowFreq}(a) is the Umklapp reversal effect for a line source placed at the centre of the array. In Fig.~\ref{fig:LowFreq}(b) is another array with the same number of units cells, but comprising only the geometry of region 1, i.e. there is no abrupt transition region to promote Umklapp scattering. Instead one observes excitation of a surface wave which is not reverse converted, and no focal spots are seen. We note that the position of the focal spots in Fig.~\ref{fig:FullDisp} is different from that of Fig.~1(a), since the angle of conversion is frequency dependent.
 We have also studied the reversed conversion effect for lower permittivity materials also with similar success. The analysis presented above is independent of the permittivity; all that is required is the inspection of the dispersion curves and overlap of similar modal shapes in the regions of differing periodicity. Higher permittivity devices are preferential for focusing applications since there is less interference from the source excitation as compared to lower permittivity materials. For optimal performance the transition regions have to be designed far enough away from the source as to not be masked by the excitation, but near enough so that the focus of the effect is still visible.

 For clarity we also present the full dispersion curves for regions 1 and 2 in Fig.~3, shown separately in Fig.~\ref{fig:FullDisp}. These curves are obtained using the Comsol multiphysics FEM software.

\end{document}